\documentclass[
reprint,
 amsmath,amssymb,
 aps,
]{revtex4-2}

\usepackage{graphicx}
\usepackage{dcolumn}
\usepackage{bm}
\usepackage{xcolor}
\usepackage{soul}
\begin{document}

\preprint{APS/123-QED}
\title{Nonlinearity-Selective Quasi–Bound States in the Continuum via Symmetry-Protected Decoupling in $\chi^{(2)}$ Thin Films}

\author{Ardra Muriyankandathil}

\author{Parikshit Sahatiya}

\author{Brian Abbey}

\author{Nitish Kumar Gupta}
\email{nitishkumar.gupta@hyderabad.bits-pilani.ac.in}

\affiliation{Department of Electrical \& Electronics Engineering,
Birla Institute of Technology and Science (BITS) Pilani,
Hyderabad Campus, India}
\affiliation{Mathematical and Physical Sciences,
La Trobe University, Australia}

\date{\today}

\begin{abstract}

Second-harmonic generation in resonant structures is commonly evaluated in terms of intracavity field enhancement at the fundamental and harmonic frequencies. Here, we formulate nonlinear frequency conversion within a symmetry-resolved overlap framework that explicitly separates resonant field buildup from nonlinear mode projection. Using a simple and analytically tractable Fabry--Perot thin-film-on-substrate geometry, we show that, even in the presence of spectrally bright resonances at both $\omega$ and $2\omega$, the emitted second-harmonic signal can be strongly suppressed when the spatial parity of the pump-induced nonlinear polarization is incompatible with that of the radiating $2\omega$ standing-wave mode. This mechanism gives rise to nonlinearity-selective quasi-bound states in the continuum. Beyond providing a compact interpretation of these nonlinear dark states, the framework unifies pump enhancement, harmonic enhancement, and symmetry-controlled modal overlap within a single predictive metric. More broadly, it identifies thickness regimes in which resonant buildup is accompanied by constructive nonlinear coupling, and distinguishes them from regimes in which apparently favorable resonance conditions remain conversion-inactive because the nonlinear source is orthogonal to the radiating harmonic mode.
\end{abstract}
\maketitle
\section{Introduction}
Second harmonic generation (SHG) mediated by a second-order nonlinear susceptibility $\chi^{(2)}$ underpins frequency conversion across classical and quantum optics, from ultrafast pulse synthesis and spectroscopy to on-chip frequency conversion and quantum-state preparation   \cite{shen1989surface,bloembergen1968reflection,Sunetal2009,guo2016conversion,geng2024nonlinear,Mckennnetal2021, larciprete2015second, yuan2021strongly, lenzini2018integrated, borsa1997measurement, valligatla2015optical}. Despite decades of development, achieving efficient SHG in compact photonic platforms  \cite{aghigh2023shg,campagnola2011shg,chen2012shg, rao2021non, li2024enhanced, li2022efficient} remains challenging, particularly when the interaction volume is reduced to wavelength/subwavelength dimensions. At this scale, nonlinear conversion is no longer governed solely by material properties or interaction length, but by how optical resonances, symmetry, and interference collectively shape the coupling between the pump field and the generated harmonic. 

In bulk nonlinear crystals, coherent build-up of the second-harmonic field is enforced through phase matching; material dispersion otherwise causes dephasing that limits the effective interaction length \cite{shen1984principles,boyd2003nonlinear}. On the other hand, at deep subwavelength scales, plasmonic nanostructures can enhance local fields, but absorption, heating, and strong material dispersion complicate both efficiency and physical interpretation, while the active nonlinear volume remains small \cite{zhang2024plasmon,noor2020mode}. In this context, high-index dielectric resonators and photonic-crystal cavities have emerged as an important alternative, offering substantially reduced dissipative loss together with the possibility of exceptionally high quality factors \cite{soljacic2002optimal,meade2008photonic,yanik2003high,smith1970intracavity,liscidini2006second,minkov2019doubly,malvezzi2003resonant}, particularly when their modal structure is engineered through symmetry-protected or accidental bound states in the continuum (BICs) \cite{cong2019symmetry, han2019all, yezekyan2024engineering}. Large Q-factors can strongly enhance intracavity fields and are often used as a proxy for nonlinear efficiency. However, nonlinear frequency conversion cannot be predicted solely from Q-factors, and our work testifies that, indeeed, this expectation is not, by itself, a reliable design principle. Resonant build-up is a necessary ingredient, but it is not sufficient, because the conversion efficiency ultimately depends on whether the pump-induced nonlinear polarization constitutes a symmetry- and phase-compatible drive for the radiating second-harmonic channel. When interference and symmetry conspire to render that drive orthogonal (in the appropriate overlap sense) to the available $2\omega$ mode, the contributions generated across the structure cancel globally. Consequently, a device can remain fully linearly bright and strongly resonant at both the pump and harmonic frequencies, yet exhibit a severely suppressed, or even vanishing, net SHG response because the nonlinear coupling pathway itself is symmetry-forbidden. We refer to such configurations as Nonlinear Dark States.

The physical origin of these states can be understood in terms of the spatial symmetry of the standing waves. In the planar geometry considered here, the nonlinear polarization $E_\omega^2(z)$ is necessarily symmetric about the film mid-plane, while the $2\omega$ cavity mode can be either symmetric or antisymmetric depending on thickness and dispersion. At specific thicknesses, the $2\omega$ standing wave becomes antisymmetric, causing contributions from opposite halves of the film to cancel pairwise in the nonlinear overlap integral. The resulting suppression of SHG does not arise from loss, weak confinement, or conventional phase mismatch, but from a symmetry-protected vanishing of the nonlinear coupling itself. In this sense, the system realizes a \emph{nonlinearity-selective quasi–bound state in the continuum}: it is not a state that fails to radiate linearly, but a configuration that fails to couple nonlinearly. The linear quality factor remains finite, yet the quadratic frequency-conversion pathway is symmetry-forbidden.

Guided by this physical picture, we deliberately restrict attention to the simplest nontrivial architecture (a single wavelength-scale $\chi^{(2)}$ planar film on an optically thick substrate) in order to examine the mechanism in its most transparent form. In this strictly one-dimensional interference setting, the pump and harmonic fields inside the film admit closed-form Fabry–Pérot standing-wave representations at $\omega$ and $2\omega$. We then introduce a spatial-mode overlap coefficient $\beta(d)$, defined as the Cauchy–Schwarz–normalized inner product between the distributed nonlinear source $E_\omega^2(z)$ and the radiating $2\omega$ cavity mode. This construction cleanly factors out mere intracavity build-up and isolates the geometric content of the $\chi^{(2)}$ interaction (phase, parity, and sign compatibility across the film thickness), thereby providing a quantitative diagnostic for nonlinearity-selective quasi-BICs in which linear resonances persist while the nonlinear channel is symmetry-suppressed. Within this context, thickness tuning is not simply resonance matching it acts as a symmetry control knob that can deterministically switch quadratic emission on or off without modifying the underlying geometry. Consequently, a key conceptual distinction emerging from our study is that cavity-enhanced nonlinear conversion is governed not only by resonance strength, but also by the spatial compatibility between the nonlinear source and the radiating harmonic mode. Thicknesses that appear optimal under purely linear or $Q$-based criteria can, in fact, proclaim a subdued nonlinear efficiency due to symmetry-imposed cancellation. By explicitly separating these effects, our framework clarifies the physical origin of nonlinear dark states and establishes spatial-mode correlation as the appropriate figure of merit for interference-engineered frequency conversion.

In the examples below, we consider a pump near $1.55~\mu$m (the telecommunications C-band, widely adopted in low-loss fiber links and silicon-compatible photonic integrated circuits \cite{lu2019entangled}) and its second harmonic near $0.78~\mu$m, frequently used in quantum photonics experiments. This $\omega$ to $2\omega$ spectral translation provides a concrete application-level motivation for studying thickness-controlled $\chi^{(2)}$ conversion in compact planar platforms where the same interference and symmetry-governed overlap physics that we highlight here directly constrains whether a device can act as a viable bridge between mature telecom-wavelength photonics and shorter-wavelength quantum photonic interfaces, even when the structure is strongly resonant at both $\omega$ and $2\omega$. While this wavelength-level motivation provides a useful application context, the central focus of this work is not the specifics of any single material system, but rather the universal interference- and symmetry-controlled mechanisms that govern nonlinear mode coupling across platforms. 

Finally, before proceeding into the details, we note that in the discussion that follows we will be using the terms nonlinearity-selective quasi-BICs and nonlinear dark states interchangeably, as both refer to the same thickness-tuned class of configurations; the latter term is adopted when emphasis is placed on the suppression of SHG despite pronounced linear resonances.

\section{Planar thin-film model: interference picture}

We consider perhaps the simplest possible nanophotonic platform, schematically represented in the inset of Fig. 1: thin--film--on--substrate with Medium 1 (air, index $n_1$), Medium 2 (a nonlinear film of thickness $d$, with indices $n_2$ at $\omega$ and $n_{2\omega}$ at $2\omega$), and Medium 3 (an optically thick substrate, index $n_3$). The structure is driven from air by a normally incident monochromatic plane wave $E_i(\omega)$ at vacuum wavelength $\lambda_0$. Because the illumination and geometry are both one-dimensional, all relevant physics is captured by the longitudinal field envelopes $E_\omega(z)$ and $E_{2\omega}(z)$ inside the film, with $z \in [0,d]$.

\subsection{Pump field as a self-consistent cavity mode}

At the pump frequency $\omega$, the electric field inside the nonlinear film satisfies the source-free scalar Helmholtz equation
\begin{equation}
 \frac{d^2 E_\omega(z)}{dz^2} + k_\omega^2 E_\omega(z) = 0,
 \qquad
 k_\omega = \frac{2\pi n_2}{\lambda_0},
 \label{eq:Helmholtz_omega}
\end{equation}
with boundary conditions at $z=0$ and $z=d$ fixed by continuity of the tangential field components across the film interfaces. The general solution of Eq.~\eqref{eq:Helmholtz_omega} inside the slab can therefore be written as a superposition of forward- and backward-propagating waves,
\begin{equation}
 E_\omega(z) = A_\omega\, e^{i k_\omega z} + B_\omega\, e^{-i k_\omega z},
 \label{eq:Eomega_general}
\end{equation}
where the amplitudes $A_\omega$ and $B_\omega$ are uniquely fixed by matching the internal field to the external plane waves in Media 1 and 3. For a systematic elimination of these intermediate amplitudes, it is convenient to encode propagation through the film in a $2\times2$ transfer matrix,
\begin{equation}
\begin{aligned}
 \begin{pmatrix}
  E_\omega(0^-)\\
  E'_\omega(0^-)
 \end{pmatrix}
 &=
 M_\omega(d)
 \begin{pmatrix}
  E_\omega(d^+)\\
  E'_\omega(d^+)
 \end{pmatrix},\\
 M_\omega(d) &=
 \begin{pmatrix}
  \cos k_\omega d & \dfrac{1}{k_\omega}\sin k_\omega d\\
  -k_\omega \sin k_\omega d & \cos k_\omega d
 \end{pmatrix}
\end{aligned}
\end{equation}
which compactly captures all phase accumulation due to scalar wave propagation within the homogeneous nonlinear film. This propagation matrix is combined with the boundary effects associated with dielectric discontinuities at $z=0$ and $z=d$ (in terms of standard Fresnel reflection and transmission coefficients at normal incidence, $r_{ij}=\frac{n_i-n_j}{n_i+n_j}$ and $t_{ij}=\frac{2n_i}{n_i+n_j}$, for $(i,j)\in\{1,2,3\}$, evaluated at frequency $\omega$). Eliminating the internal amplitudes in this way yields a closed-form expression for the intracavity field in which all multiple reflections/round-trips are summed (infinite summation) into a single Fabry-Perot denominator (further details in supporting information file): 
\begin{equation}
 D_\omega(d) = 1 - r_{12} r_{23}\, e^{2 i k_\omega d}
 \label{eq:Domega}
\end{equation}
The longitudinal pump field inside the film can then be written in the compact interference form
\begin{equation}
 E_\omega(z)
 =
 \frac{t_{12} E_i}{D_\omega(d)}
 \left[
  e^{i k_\omega z}
  +
  r_{23} e^{i k_\omega (2d - z)}
 \right]
 \qquad 0 \le z \le d
 \label{eq:Eomega_cavity}
\end{equation}
Equation~\eqref{eq:Eomega_cavity} makes the self-consistent cavity character explicit: the denominator $D_\omega(d)$ accounts for multiple reflections within the film, while the numerator describes the interference between the two partial waves generated at the boundaries. Thicknesses for which $|D_\omega(d)|$ is minimized correspond to pump resonances with strong standing-wave build-up.

\subsection{Second-harmonic field as a driven cavity mode}

At the second-harmonic frequency $2\omega$, the field inside the nonlinear film is driven by the distributed nonlinear polarization induced by the pump and is simultaneously constrained by the same cavity boundaries. The scalar Helmholtz equation now reads
\begin{equation}
\begin{aligned}
   \frac{d^2 E_{2\omega}(z)}{dz^2} + k_{2\omega}^2 E_{2\omega}(z)
 =
 -\,\mu_0 (2\omega)^2 P^{(2)}_{2\omega}(z)\\
 \qquad
 k_{2\omega} = \frac{n_{2\omega}\,2\omega}{c}
 \quad
 \label{eq:Helmholtz_2omega}
\end{aligned}
\end{equation}  
with a nonlinear source term
\begin{equation}
 P^{(2)}_{2\omega}(z)
 = \varepsilon_0 \chi^{(2)} E_\omega^2(z).
 \label{eq:Pnl}
\end{equation}

To expose the interference structure, we express the solution of Eq.~\eqref{eq:Helmholtz_2omega} in terms of the scalar Green function $G_{2\omega}(z,z';d)$ of the linear cavity at $2\omega$,
\begin{equation}
 \left[
  \frac{d^2}{dz^2} + k_{2\omega}^2
 \right]
 G_{2\omega}(z,z';d)
 = \delta(z - z')
 \label{eq:Green_def}
\end{equation}
subject to the same boundary conditions at $z=0$ and $z=d$ as the physical field $E_{2\omega}(z)$. A convenient representation that keeps the cavity physics explicit, yet remains interpretable, is obtained by writing the $2\omega$ Green function as a direct propagator plus its first-reflection images, dressed by the Fabry--Perot round-trip series:
\begin{equation}
\begin{split}
 G_{2\omega}(z,z';d) = \frac{i}{2 k_{2\omega} D_{2\omega}(d)} \Bigl[ &e^{i k_{2\omega}|z-z'|} 
 + r^{(2\omega)}_{12}\, e^{i k_{2\omega}(z+z')} \\
 &+ r^{(2\omega)}_{23}\, e^{i k_{2\omega}(2d-z-z')} \\
 &+ r^{(2\omega)}_{12} r^{(2\omega)}_{23}\, e^{i k_{2\omega}(2d-|z-z'|)} \Bigr]
\end{split}
\label{eq:Green_explicit}
\end{equation}
Here, the four numerator terms correspond, respectively, to direct propagation from $z'$ to $z$, reflection at $z=0$, reflection at $z=d$, and a round-trip contribution; higher-order multiple reflections are absorbed into the Fabry-Perot denominator
\begin{equation}
 D_{2\omega}(d) = 1 - r^{(2\omega)}_{12} r^{(2\omega)}_{23}\, e^{2 i k_{2\omega} d}
 \label{eq:D2omega}
\end{equation}
with $r^{(2\omega)}_{12}$ and $r^{(2\omega)}_{23}$ the normal-incidence reflection coefficients evaluated at $2\omega$. The particular solution of Eq.~\eqref{eq:Helmholtz_2omega} inside the film can then be written in Green-function form as
\begin{equation}
 E_{2\omega}(z)
 =
 -\,\mu_0 (2\omega)^2
 \int_0^d
 G_{2\omega}(z,z';d)\,
 P^{(2)}_{2\omega}(z')\, dz'
 \label{eq:E2omega_green}
\end{equation}
which expresses the generated second-harmonic field as the linear cavity response at $2\omega$ to a distributed nonlinear source. Substituting Eq.~\eqref{eq:Pnl} shows that the spatial structure of $E_{2\omega}(z)$ is determined by how the pump-induced polarization $E_\omega^2(z')$ is resolved by the Green operator of the Fabry-Perot cavity. Because the cavity at $2\omega$ is linear, its Green function admits a representation in terms of its standing-wave eigenmodes; consequently, the spatial dependence relevant for nonlinear coupling is entirely captured by the corresponding cavity mode profile, while the source strength enters only as a scalar weight. For the spatial-mode-overlap and enhancement-factor analysis in Sec. 3, it is therefore sufficient to factor out this source-dependent amplitude and retain the normalized longitudinal mode shape at $2\omega$. Carrying out this reduction yields
\begin{equation}
 E_{2\omega}(z)
 =
 \frac{t^{(2\omega)}_{12}}{D_{2\omega}(d)}
 \left[
  e^{i k_{2\omega} z}
  +
  r^{(2\omega)}_{23} e^{i k_{2\omega}(2d - z)}
 \right]
 \qquad 0 \le z \le d
 \label{eq:E2omega_mode}
\end{equation}
where all multiple reflections and phase accumulation are absorbed into the Fabry-Perot denominator $D_{2\omega}(d)$.

At this stage, the longitudinal field distributions at $\omega$ and $2\omega$ are fully specified by Eqs.~\eqref{eq:Eomega_cavity} and~\eqref{eq:E2omega_mode} (more details in supporting information file). The denominators $D_\omega(d)$ and $D_{2\omega}(d)$ in these equations emerge as the central quantities governing double resonance \cite{ou1993enhanced}. The remaining question is therefore not one of field existence or resonant build-up, but of how efficiently the pump-induced nonlinear polarization couples to the radiating harmonic mode through their spatial correlation across the film thickness.

\section{Spatial-mode overlap and enhancement factor}
\label{sec:beta}

\subsection{Normalized overlap coefficient}

The SHG efficiency in any inhomogeneous structure depends not only on the strength of modal confinement, but also, more fundamentally, on the constructive spatial overlap between the pump field $E_{\omega}(z)$ and the radiating second-harmonic field $E_{2\omega}(z)$. In our planar film, this overlap can be captured by a single dimensionless quantity, \cite{lin2016cavity,rodriguez2007chi2,noor2020mode} which we call the normalized nonlinear coupling coefficient $\beta(d)$, defined as
\begin{equation}
\beta(d) =
\frac{\displaystyle\int_0^d E_\omega^2(z)\, E_{2\omega}^\ast(z)\, dz}
{\sqrt{\left(\displaystyle\int_0^d |E_\omega(z)|^4 dz\right)
\left(\displaystyle\int_0^d |E_{2\omega}(z)|^2 dz\right)}}
\label{eq:beta}
\end{equation}
Introducing $F(z) = E_{\omega}^2(z)$ and $G(z) = E_{2\omega}(z)$, we may write Eq.~\eqref{eq:beta} in the compact inner-product form
\begin{equation}
\beta(d) = 
\frac{\langle F, G \rangle}
{\sqrt{\|F\|^{2}\, \|G\|^{2}}}
\end{equation}
with $\langle F,G\rangle=\int_0^d F(z)\,G^\ast(z)\,dz$ and $\|F\|^2=\int_0^d |F(z)|^2 dz$. By the Cauchy--Schwarz inequality, $|\beta(d)| \le 1$ for all $d$ (further details in supporting information file)
                                \begin{figure}[t!]
                                  \centering{\includegraphics[width=\linewidth]{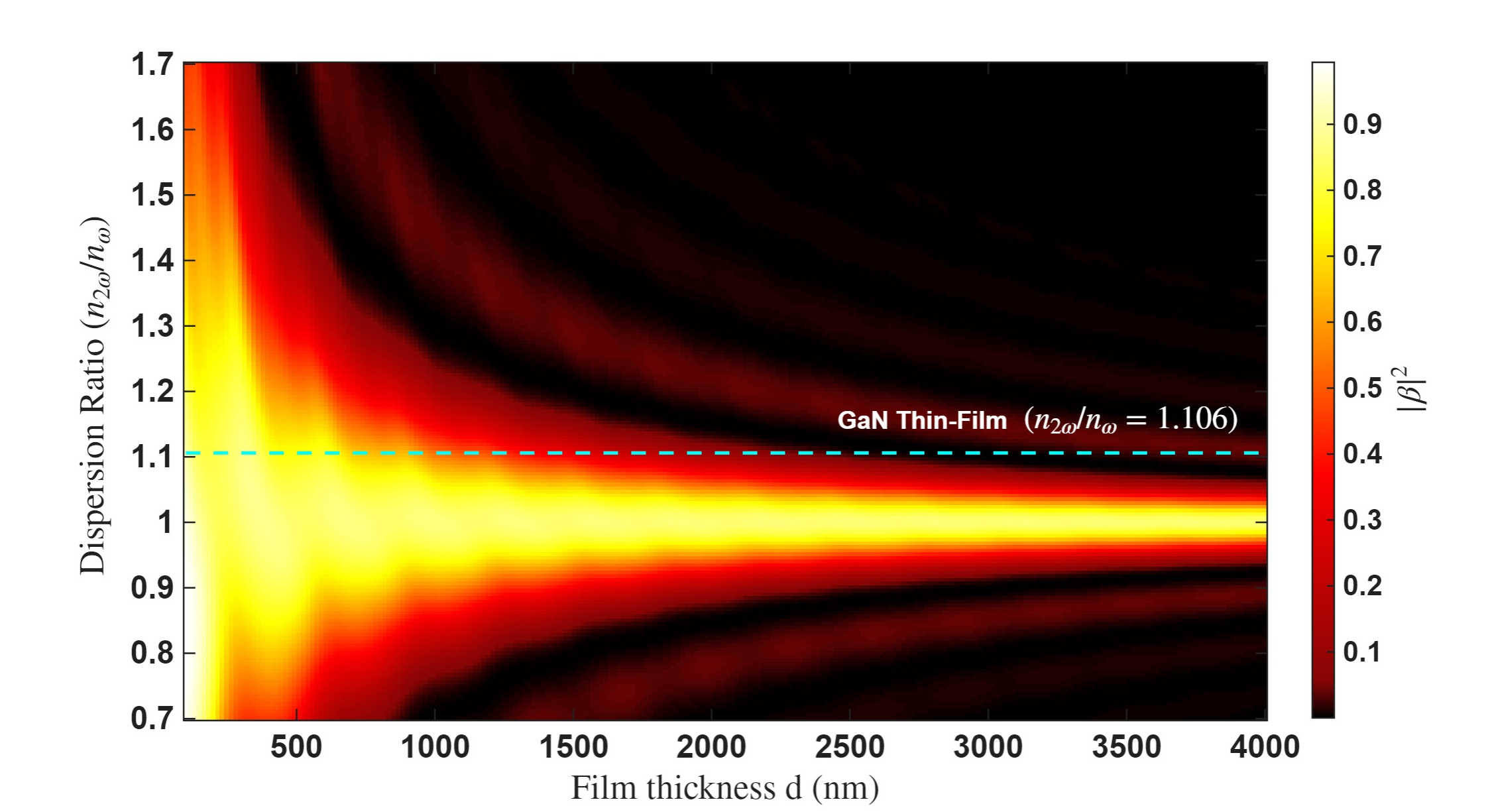}}
                                 \caption{Thickness--dispersion map of the normalized nonlinear overlap coefficient $|\beta(d)|^{2}$ [Eq. \eqref{eq:beta}] in the $(d,n_{2\omega}/n_{\omega})$ plane. Bright ridges identify thickness--dispersion combinations for which the pump-induced nonlinear polarization $E_{\omega}^{2}(z)$ is symmetry-compatible with the radiating $2\omega$ standing wave.}
                                 \label{fig:beta_heatmap}
                                \end{figure}
The normalization of $\beta(d)$ ensures that $|\beta| \leq 1$, with the upper bound attained only when the pump-induced nonlinear polarization $E_\omega^2(z)$ is everywhere phase- and symmetry-matched to the radiating second-harmonic eigenmode $E_{2\omega}(z)$. In that limit, contributions from all slices of the film add constructively, and the nonlinear coupling is maximized in the Cauchy--Schwarz sense. Viewed this way, the longitudinal field distributions, boundary conditions, and interference effects at both frequencies are compactly encoded into a single scalar function $\beta(d)$, which depends solely on the film thickness and material dispersion. This reduction makes $\beta(d)$ a natural tool for identifying regimes of constructive and destructive nonlinear interference. Accordingly, maps of $|\beta(d)|$ in the $(d,\,n_{2\omega}/n_2)$ parameter space provide a direct visualization of where nonlinear contributions reinforce or cancel across the film thickness. Fig. 1 makes it explicit that the overlap landscape is highly structured. Bright ridges identify parameter combinations for which the nonlinear polarization $E_\omega^{2}(z)$ and the radiating $2\omega$ standing wave remain globally phase-compatible across the film, whereas the narrow dark valleys correspond to thicknesses where the overlap is suppressed by symmetry-driven cancellation. Importantly, these valleys do not signal a disappearance of the underlying linear cavity fields; rather, they mark points where the nonlinear coupling channel is quenched even though the film can remain resonant and radiatively bright at the relevant frequencies. These features of $\beta(d)$ landscape underscore the central message of our work: cavity build-up alone is not a sufficient metric for SHG performance, and the overlap coefficient $\beta(d)$ is the decisive ingredient that distinguishes genuinely phase-coherent, interference-engineered thicknesses from those that are nearly superficially resonant. 

\subsection{Enhancement factor and decomposition}

In the undepleted-pump regime, the total radiated second-harmonic power can be written as
\begin{equation}
W(2\omega) \propto |\chi^{(2)}|^{2}
\left| \int_{0}^{d} E_{\omega}^{2}(z)\, E_{2\omega}^{*}(z)\, dz \right|^{2}
\label{eq:W2wProportional}
\end{equation}

                        \begin{figure*}[t!]
                        \centering{\includegraphics[width=\textwidth]{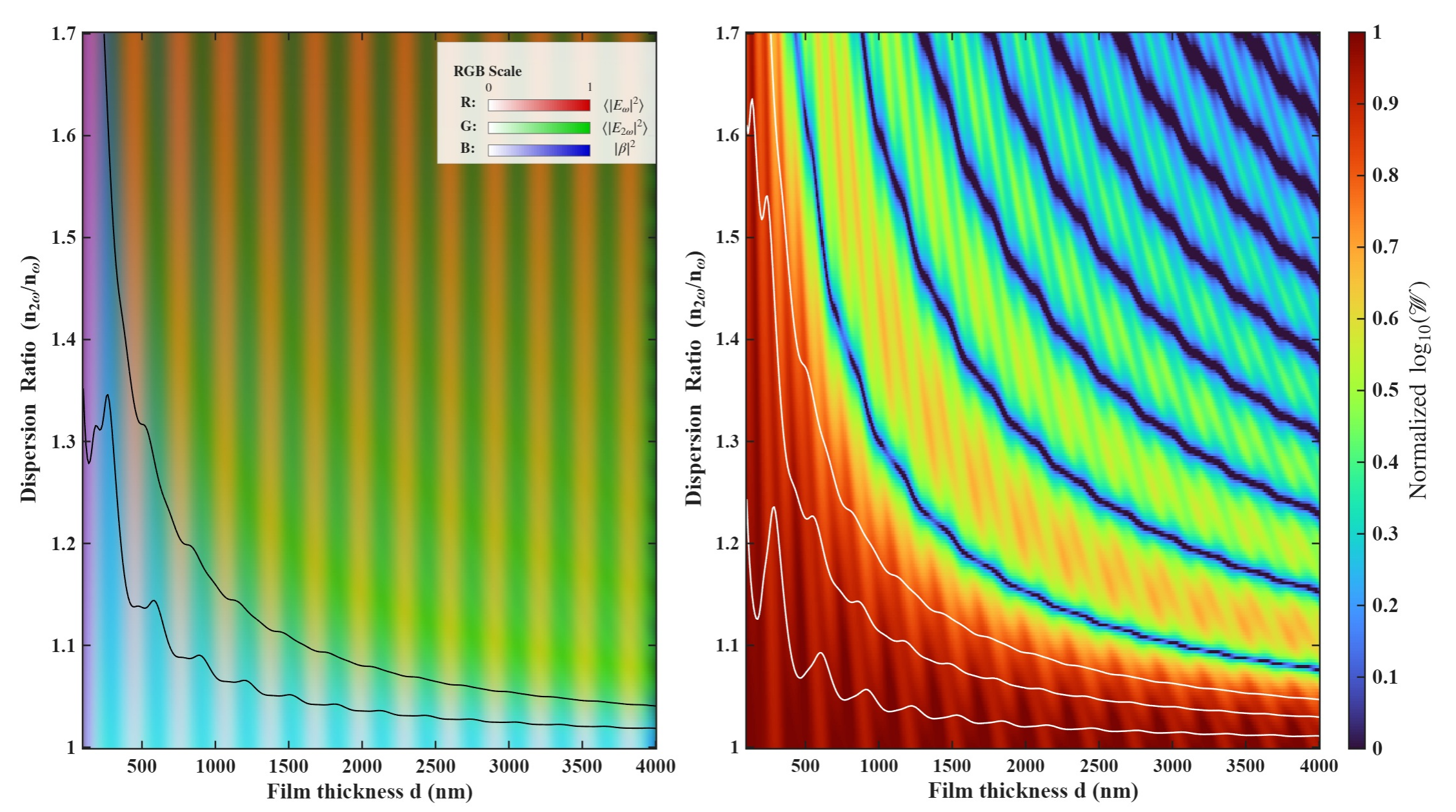}}
                        \caption{Design maps for doubly resonant $\chi^{(2)}$ thin films:
                        (a) An RGB illustrative map in the $(d,\,n_{2\omega}/n_{\omega})$ plane. The red and green channels encode the depth-averaged intracavity build-up $\langle|E_{\omega}|^2\rangle_d$ and $\langle|E_{2\omega}|^2\rangle_d$, respectively, while the blue channel encodes the normalized nonlinear overlap $|\beta(d,n_{2\omega}/n_{\omega})|^2$. (b) Normalized logarithmic map of the optical overlap--buildup merit function $\mathcal{W}$ defined from Eq.~17 of the manuscript by isolating its field-dependent contribution. Contours indicate representative levels of $|\beta|^2$, highlighting overlap-limited valleys where large linear build-up does not translate into strong conversion.}
                        \end{figure*}
For later convenience, we denote the overlap integral by $I(d) = \int_0^d E_\omega^2(z)\,E_{2\omega}^\ast(z)\,dz$. Using the definition of $\beta(d)$ in Eq.~\eqref{eq:beta}, its magnitude factorizes as
\begin{equation}
 |I(d)|
 = |\beta(d)|
   \left[
     \int_0^d |E_\omega(z)|^4\, dz
     \int_0^d |E_{2\omega}(z)|^2\, dz
   \right]^{1/2}
 \label{eq:I_factorized}
\end{equation}
so that
\begin{equation}
 W_{2\omega}(d)
 \propto
 |\chi^{(2)}|^2
 \, d^2
 \,
 \big\langle |E_\omega|^2 \big\rangle_d^{\,2}
 \,
 \big\langle |E_{2\omega}|^2 \big\rangle_d
 \,
 |\beta(d)|^2
 \label{eq:W_factorized}
\end{equation}
where the depth-averaged intensities are $\big\langle |E_\omega|^2 \big\rangle_d = \frac{1}{d} \int_0^d |E_\omega(z)|^2\, dz$ and $\big\langle |E_{2\omega}|^2 \big\rangle_d = \frac{1}{d} \int_0^d |E_{2\omega}(z)|^2\, dz$. Equation~\eqref{eq:W_factorized} cleanly separates the roles of (i) pump-field build-up, (ii) second-harmonic build-up and out-coupling, and (iii) the purely geometric interference encoded in $|\beta(d)|^2$.
We observe here that equation~17 separates naturally into a material/geometric prefactor (containing, e.g., $\chi^{(2)}$ and film thickness) and a purely optical contribution governed by intracavity build-up at $\omega$ and $2\omega$ and by the overlap penalty $|\beta|^2$. To visualize this optical part across thickness and dispersion, we define an overlap--buildup merit function by retaining only the field-dependent factor in Eq.~17, namely
$\mathcal{W}(d,n_{2\omega}/n_{\omega}) \equiv \langle|E_{\omega}|^2\rangle_d^{2}\,\langle|E_{2\omega}|^2\rangle_d\,|\beta(d,n_{2\omega}/n_{\omega})|^2$,
where $\langle\cdot\rangle_d$ denotes depth-averaging across the film.
In Fig.2(b), we plot the normalized logarithmic map of this quantity, $\mathrm{Norm}\{\log_{10}\mathcal{W}\}$, so that the wide dynamic range of cavity enhancement is represented compactly. Fig.2(a) provides an illustrative decomposition by displaying $\langle|E_{\omega}|^2\rangle_d$, $\langle|E_{2\omega}|^2\rangle_d$, and $|\beta|^2$ as RGB channels, making explicit when doubly resonant build-up is rendered ineffective by overlap-limited valleys.

\noindent Furthermore, to illustrate the physical utility of this framework, we benchmark our minimal thin-film structure against a bulk medium: ( more details in supporting information file)
\begin{equation}
W_{\mathrm{bulk}}(2\omega)
\propto
|\chi^{(2)}|^{2}\, d^{2}\,
|E_{\omega,\mathrm{bulk}}|^{4}\,
|E_{2\omega,\mathrm{bulk}}|^{2}\,
|\beta_{bulk}|^2
\label{eq:w_bulk}
\end{equation}
where $E_{\omega,\mathrm{bulk}}$, $E_{2\omega,\mathrm{bulk}}$ the internal plane-wave amplitudes. The corresponding enhancement factor is then defined as
\begin{equation}
\mathcal{E}(d)
=
\frac{W(2\omega)}{W_{\mathrm{bulk}}(2\omega)}
=
\left[
\frac{\big\langle |E_{\omega}|^{2} \big\rangle_d}{\left| E_{\omega,\mathrm{bulk}} \right|^{2}}
\right]^{2}
\,
\left[
\frac{\big\langle |E_{2\omega}|^{2}\big\rangle_d}{\left| E_{2\omega,\mathrm{bulk}} \right|^{2}}
\right]
\,
\left[
\frac{|\beta(d)|}{|\beta_{\mathrm{bulk}}|}
\right]^{2}
\label{eq:EnhancementFactor}
\end{equation}
Here, the first two brackets play the role of effective $Q$-like build-up factors at $\omega$ and $2\omega$, respectively, while the third bracket isolates the purely geometric overlap penalty or gain relative to the bulk reference. In a homogeneous slab where $\Delta k = 2k_\omega - k_{2\omega} = 0$ (ideal, perfectly phase-matched bulk) over an effectively infinite coherence length, one finds $|\beta_{\mathrm{bulk}}| = 1$, and the enhancement factor simplifies to
\begin{equation}
\mathcal{E}(d)
=
\frac{W(2\omega)}{W_{\mathrm{bulk}}(2\omega)}
=
\left[
\frac{\big\langle |E_{\omega}|^{2} \big\rangle_d}{\left| E_{\omega,\mathrm{bulk}} \right|^{2}}
\right]^{2}
\,
\left[
\frac{\big\langle |E_{2\omega}|^{2}\big\rangle_d}{\left| E_{2\omega,\mathrm{bulk}} \right|^{2}}
\right]
\,
\left[
\frac{|\beta(d)|}{1}
\right]^{2}
\label{eq:EnhancementFactor_Simplified}
\end{equation}

Guided by this component-wise decomposition of the enhancement factor, we now embark to examine the thickness dependence of  $\mathcal{E}(d)$, which will help us elucidate the underlying trends. However, before analyzing Eq.~\eqref{eq:EnhancementFactor_Simplified} in its full form, it is instructive to adopt the incomplete, yet notionally common, viewpoint in which the overlap term is tacitly assumed ideal. Particularly, setting $|\beta(d)| \equiv 1$ in Eq.~\eqref{eq:EnhancementFactor_Simplified} removes the third bracket and yields an ``apparent'' enhancement coefficient $\mathcal{E}_{\mathrm{app}}(d)$ that depends only on the effective cavity build-up factors at $\omega$ and $2\omega$.
Fig. 3 shows $\mathcal{E}_{\mathrm{app}}(d)$ as a function of thickness for a representative material system: GaN (refractive index $n \approx 2.55$ at 1560 nm) on sapphire substrate \cite{zaky2025gan,roland2016gan,miragliotta1993linear,abe2010accurate,yu1997gan}. The choice of GaN-on-sapphire here is not meant as an experimental constraint, but as a concrete and accepted, wide-bandgap $\chi^{(2)}$ thin-film system that is simultaneously relevant to near-infrared photonics and to near-visible harmonic wavelengths. This application-level framing is intentionally kept out of discussion till now as our central claims do not rely on any GaN-specific fabrication pathway, and the symmetry-controlled overlap mechanism developed in Eqs. (13)--(20) applies generically to any dispersive $\chi^{(2)}$ film-on-substrate platform once $(n_{\omega},n_{2\omega})$ and the boundary conditions are specified.

\begin{figure}[b!]
 \centering{\includegraphics[width=\linewidth]{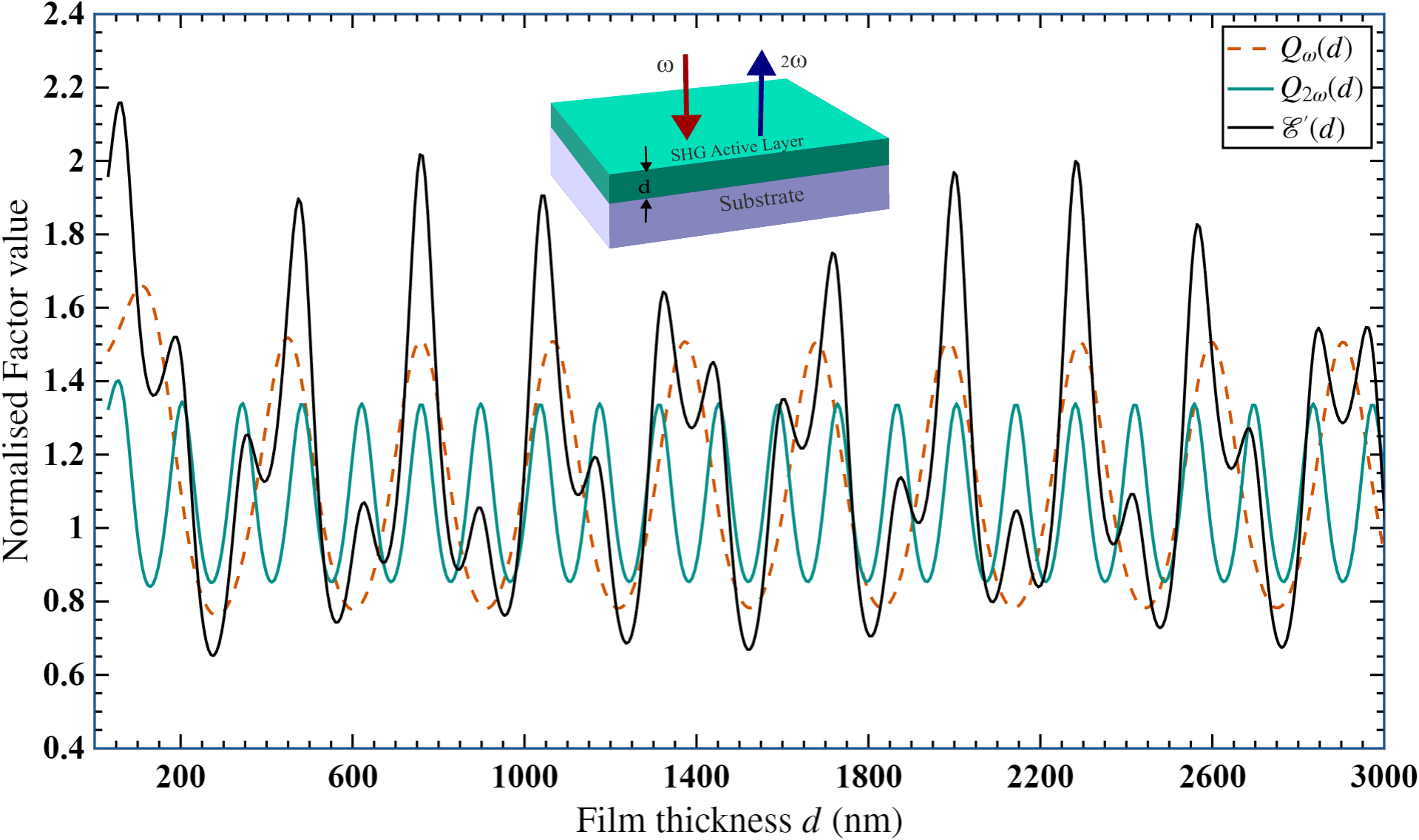}}
 \caption{Apparent enhancement factor $\mathcal{E}_{\mathrm{app}}(d)$ that includes only the cavity build-up factors and explicitly neglects the $|\beta(d)|^{2}$ term by setting $|\beta(d)| \equiv 1$ in Eq.~\eqref{eq:EnhancementFactor}. This is representative of a conventional $Q$-driven cavity-enhanced SHG design.}
 \label{fig:Eapp}
\end{figure}
With this physically motivated material and wavelength choice fixed, we now return to the central question, namely, how the thickness-dependent field interference and mode-overlap encoded in $\mathcal{E}(d)$ manifest themselves in a planar system (by examining the calculated enhancement shown in Fig. 3). The curve exhibits a series of prominent peaks whenever the Fabry--Pérot conditions at the two frequencies are approximately satisfied, and would, on its own, suggest that very large SHG enhancements are available simply by tuning $d$ to align these resonances. This is essentially the notional design logic in cavity-enhanced SHG, where one predominantly focuses on maximizing $Q$-like intracavity intensities (implicitly assuming that the nonlinear polarization and the radiating $2\omega$ mode are perfectly matched in space and phase) \cite{koshelev2020subwavelength,carletti2018giant,https://doi.org/10.1002/adpr.202400167}.

However, Eq.~\eqref{eq:EnhancementFactor_Simplified} makes clear that this picture is incomplete: the third bracket, $|\beta(d)|^{2}$, measures the actual spatial-mode correlation inside the nonlinear film, and any deviation of $|\beta(d)|$ from unity multiplicatively suppresses the idealized enhancement. When this geometric factor is restored, the full enhancement $\mathcal{E}(d)$ in Fig. 4 no longer follows the optimistic values suggested by $\mathcal{E}_{\mathrm{app}}(d)$ of Fig. 3: in fact the entire family of apparent resonant peaks are pulled down because large intracavity fields are accompanied by strong internal cancellation in the overlap integral. The contrast between Figs. 3 and 4 reiterates the importance of overlap coefficient $\beta(d)$ in identifying the most favorable regime for frequency conversion. 

\begin{figure}[htbp]
 \centering{\includegraphics[width=\linewidth]{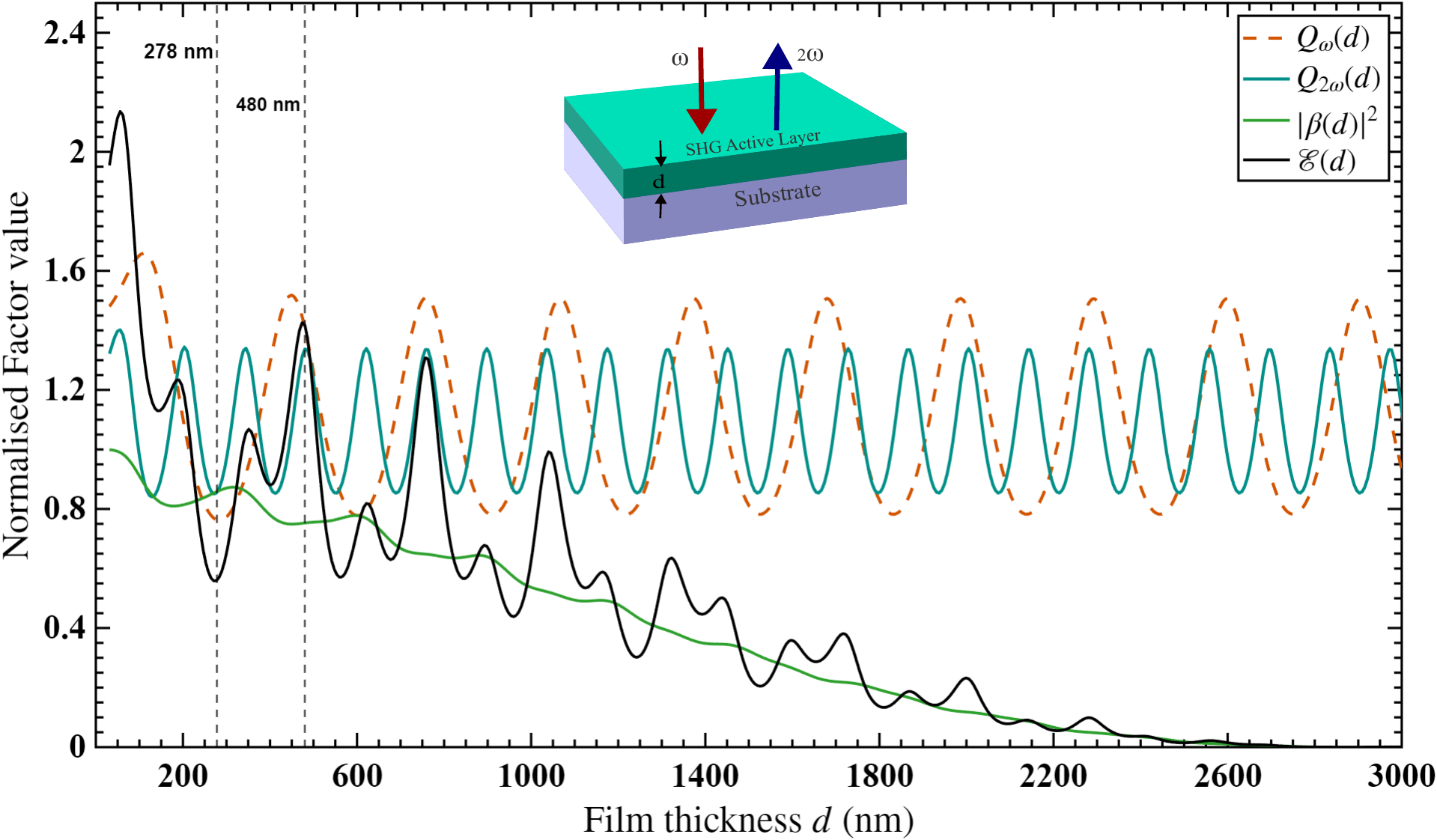}}
 \caption{Enhancement factor $\mathcal{E}(d)$ and its constituents as a function of film thickness $d$. The solid black shows the total enhancement predicted by Eq.~\eqref{eq:EnhancementFactor}, while the orange and cyan curves represent the pump and second-harmonic build-up factors, respectively. The green curve tracks the purely geometric contribution $|\beta(d)|^{2}$. The vertical lines depict two representative thicknesses for local minima/maxima of $\mathcal{E}(d)$.}
 \label{fig:Enhancement}
\end{figure}

\section{Origin of Nonlinear Dark States: A Symmetry-protected Framework} 
\label{sec:symmetry_dark_states}

We have seen that the thickness dependence of the overlap coefficient is strongly oscillatory, with pronounced maxima and minima visible in Fig. 4. We now examine the physical origin of these features and show that the oscillatory behaviour of $|\beta(d)|^{2}$ stems from a symmetry-controlled selection rule. This mechanism leads to nonlinear dark states, where cavities at $\omega$ and $2\omega$ remain linearly bright and well formed, yet the second-harmonic output is quenched beacuse spatial overlap integral cancels by parity. This provides a direct nonlinear analogue of symmetry-protected bound states in the continuum (BICs) and illustrates a type of ``nonlinearity-selective darkness'' that is completely invisible to purely linear $Q$-factor considerations \cite{yu2019bic}.

\subsection{Parity structure of the standing waves}

For normal incidence on a symmetric film (same incident medium on both sides, or equivalently after folding the geometry about the mid-plane), the longitudinal profiles $E_\omega(z)$ and $E_{2\omega}(z)$ are solutions of a one-dimensional Helmholtz problem with reflection symmetry about the film centre $z=d/2$. It is convenient to introduce the shifted coordinate $\zeta = z - \frac{d}{2}$, so that the film occupies a symmetric interval around $\zeta=0$. In the absence of material absorption, standing-wave solutions at each frequency can be chosen to be even or odd under $\zeta \to -\zeta$:
\begin{equation}
 E_{\omega}^{(\pm)}(-\zeta) = \pm E_{\omega}^{(\pm)}(\zeta),
 \qquad
 E_{2\omega}^{(\pm)}(-\zeta) = \pm E_{2\omega}^{(\pm)}(\zeta)
 \label{eq:parity_def}
\end{equation}
where the superscripts $(+)$ and $(-)$ denote even and odd parity, respectively. In an ideal Fabry–Pérot cavity with perfectly symmetric mirrors, the longitudinal mode index $m$ (or $p$ at $2\omega$) fixes this parity in the usual way: even-$m$ modes are cosine-like and even, odd-$m$ modes are sine-like and odd. In our more realistic, partially reflecting film, the parity classification is not exact in the strict eigenfunction sense; nevertheless, near thicknesses where the cavity response is well developed, the longitudinal profiles retain a robust even/odd character about the mid-plane, which is sufficient for the symmetry argument below:

The nonlinear source term that drives SHG in Eq.~\eqref{eq:W2wProportional} involves the square of the pump field, $S(z) = E_\omega^2(z)\, E_{2\omega}^\ast(z)$. Therefore, in terms of the centred coordinate $\zeta$, the parity of $E_\omega^2$ is always even, regardless of whether $E_\omega$ itself is even or odd: $E_\omega^2(-\zeta) = E_\omega^2(\zeta)$. This is simply the statement that squaring removes any sign change. As a result, the parity of the nonlinear source $S(\zeta)$ is entirely inherited from $E_{2\omega}(\zeta)$:
\begin{equation}
 S(-\zeta) =
 \begin{cases}
  +\,S(\zeta), & E_{2\omega} \ \text{even}\\[2pt]
  -\,S(\zeta), & E_{2\omega} \ \text{odd}
 \end{cases}
\end{equation}

The numerator of the overlap coefficient can be written as $\mathcal{I}(d) = \int_{-d/2}^{d/2} S(\zeta)\, d\zeta$, so that a purely odd $2\omega$ mode immediately forces the integral to vanish:
\begin{equation}
 E_{2\omega}(-\zeta) = -E_{2\omega}(\zeta)
 \quad \Rightarrow \quad
 \mathcal{I}(d) = 0,
 \quad
 \beta(d) = 0.
 \label{eq:beta_zero_condition}
\end{equation}

Equation~\eqref{eq:beta_zero_condition} provides a purely symmetry-based route to vanishing nonlinear conversion, and helps us render the interference‑based cancellation mechanism transparent for the actual air/film/substrate geometry. For completeness, we want to mention here that the symmetry arguments introduced here are formulated using an idealized mirror-symmetric film as a conceptual reference. and they do not rely on weak fields, loss, or detuning, but follows from the sign structure of the overlap kernel itself.

Finally, to expose the mechanism in real space, we introduce the local overlap density
\begin{equation}
 \rho(z,d) = \mathrm{Re}\!\left\{E_{\omega}^{2}(z;d)\,E_{2\omega}^{\ast}(z;d)\right\}
 \label{eq:rho}
\end{equation}
whose depth integral is proportional to the numerator of Eq.~\eqref{eq:beta}. Positive (negative) $\rho$ indicates regions that drive the radiating $2\omega$ channel in phase (out of phase). Fig. 5 visualizes $\rho(z,d)$ and shows that, at the dark-state thicknesses predicted by Eq.~\eqref{eq:beta_zero_condition}, $\rho(z,d)$ develops a sign-alternating structure with near-balanced positive and negative contributions, enforcing a near-zero net projection despite substantial local polarization. With these diagnostics in place, we can now reinterpret the suppression of SHG as a parity-controlled decoupling of the $\chi^{(2)}$ source from the radiating harmonic channel, and contrast it with purely $Q$-driven design principles.

\begin{figure}[htbp]
 \centering{\includegraphics[width=\linewidth]{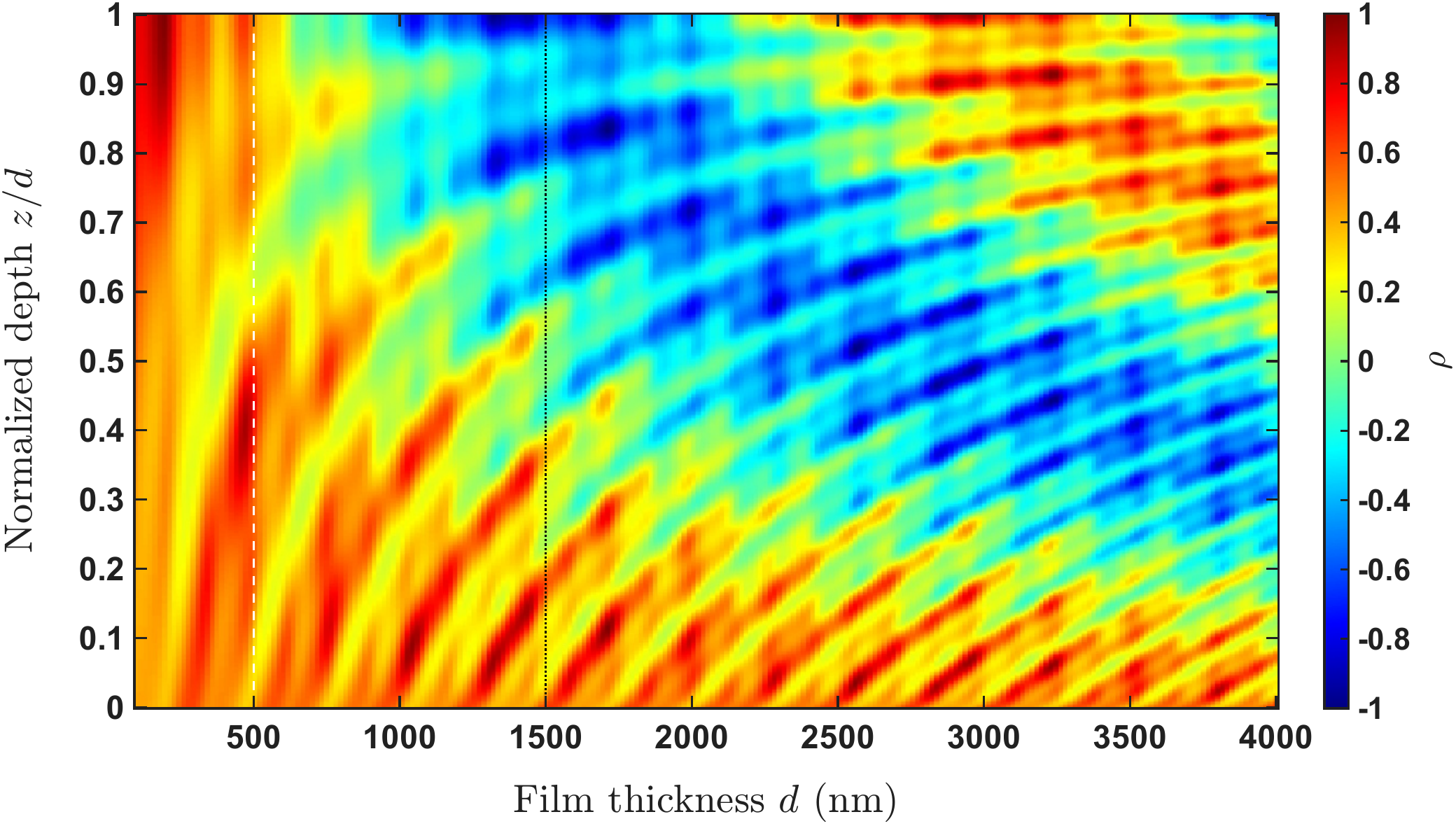}}
 \caption{Real-space structure of the nonlinear coupling channel, visualized via the normalized local overlap density $\rho(z,d)=\mathrm{Re}\{E_\omega^2(z;d)\, E_{2\omega}^\ast(z;d)\}$ [Eq. \eqref{eq:rho}], plotted versus normalized depth $z/d$ and film thickness $d$. The vertical lines depict two representative cases; one with predominantly same-sign local overlap density and another with sign-alternating local overlap density.}
 \label{fig:overlap_density}
\end{figure}

\subsection{Interpretation of Parity-controlled Nonlinear quasi-BICs and contrast with $Q$-driven design}

The condition in Eq.~\eqref{eq:beta_zero_condition} is strongly reminiscent of symmetry-protected bound states in the continuum, where a mode is completely decoupled from the radiation continuum by parity and therefore exhibits an effectively infinite linear $Q$-factor. In our planar film, by contrast, the odd $2\omega$ standing wave remains radiatively coupled to the outside world and has a finite linear $Q$; only the nonlinear coupling to the internal pump-field-induced polarization is suppressed. From the standpoint of SHG, these thicknesses behave as nonlinear BIC-like states: bright in linear optics but dark to quadratic mixing. In Fig. 2, these nonlinearity-selective quasi-BIC conditions appear as narrow valleys where $|\beta(d)|^{2}$ is suppressed, despite nearby regions of strong resonant buildup.

This distinction is particularly transparent if we compare the enhancement factor in Eq.~\eqref{eq:EnhancementFactor_Simplified} to the cavity build-up at $2\omega$. Near an odd-parity thickness, the intensity enhancement $\langle |E_{2\omega}|^2\rangle_d$ can remain substantial, signalling a well-formed resonator, while the overlap coefficient $|\beta(d)|$ passes through minima. Any design strategy that focuses solely on maximizing $Q$-factors or internal intensities can therefore select the wrong thickness from the perspective of nonlinear conversion. The interference-engineered design rule proposed here instead targets thicknesses for which the symmetry of $E_{2\omega}(\zeta)$ is compatible with the even symmetry of $E_\omega^2(\zeta)$, so that the overlap integral is driven towards its Cauchy--Schwarz bound $|\beta|\to 1$. In the numerical maps of Figs. 1 and 2, these symmetry-controlled extrema appear as valleys where $|\beta(d)|^2$ exhibits minima, separating the bright ridges associated with double-resonant, even-parity configurations. The transition from one ridge to the next necessarily passes through a nonlinear dark state where Eq.~\eqref{eq:beta_zero_condition} is approximately satisfied. This provides a concrete interpretation of the oscillatory structure of $|\beta(d)|^2$: each oscillation corresponds to a parity flip of the $2\omega$ standing wave relative to the pump-induced source distribution. The corresponding real-space cancellation is directly visible in Fig. 5, where $\rho(z,d)$ alternates in sign and integrates to nearly zero at the dark-state thicknesses.

\section{Summary \& Discussion}
In this work, we have shown that interference engineering in a single-wavelength-scale $\chi^{(2)}$ planar film is sufficient to realize a rich and underappreciated class of nonlinear states that can be overlooked within conventional design paradigms. By formulating Second harmonic generation entirely in terms of spatial-mode correlation between the pump-induced nonlinear polarization and the radiating second-harmonic cavity mode, we identified thickness-tuned configurations in which strong linear resonances coexist with vanishing nonlinear output. These configurations constitute nonlinear dark states in which the cavity is well formed and the structure is fully accessible in linear optics, yet the quadratic frequency-conversion channel is symmetry-suppressed.

Our key contribution lies in identifying a minimal and analytically tractable configuration to demonstrate that cavity-enhanced nonlinear conversion is governed not solely by resonance strength or field confinement, but by the symmetry-compatible projection of the nonlinear source onto the available radiating channel. In the symmetry reference picture used to interpret the cancellation mechanism, the nonlinear polarization $E_\omega^2(z)$ is even about the film mid-plane, whereas the $2\omega$ standing wave may be either symmetric or antisymmetric depending on thickness and dispersion. When the harmonic mode becomes antisymmetric, contributions from opposite halves of the film cancel pairwise in the overlap integral, enforcing a symmetry-protected minimum of the nonlinear coupling despite strong intracavity fields. This behavior is most naturally read through the BIC lens, but with the roles of radiation and coupling exchanged. The thicknesses singled out here do not annihilate the linear $2\omega$ radiation channel as Fabry-Perot resonances remain well defined, and the associated modes stay radiatively bright with finite linewidth. What is extinguished is the driving of that channel by $\chi^{(2)}$ mixing where the pump-induced polarization becomes symmetry-incompatible with the radiating $2\omega$ standing wave, so that the nonlinear overlap integral collapses by global cancellation. In this sense, the film realizes a nonlinearity-selective quasi-BIC: a symmetry-protected minima of the quadratic excitation pathway rather than a symmetry-protected suppression of linear leakage. The distinction matters operationally because it isolates a genuinely interference-and symmetry-imposed limitation (orthogonality between source and mode) from more familiar causes of weak SHG, such as absorption, poor confinement, or conventional phase mismatch.

Another important outcome of this work is the introduction of a properly normalized overlap coefficient $\beta(d)$, which separates cavity build-up from spatial-mode compatibility. In this role, $\beta(d)$ is not merely a figure of merit, but a diagnostic quantity for symmetry-induced nonlinear cancellation. The sharp oscillations and minima of $|\beta(d)|^2$ therefore emerge as direct signatures of parity-controlled nonlinear quasi-BICs rather than numerical artifacts or accidental interference effects.

Beyond providing an analytically transparent explanation for thickness-dependent SHG extrema in planar films, the present framework has broader implications. Thickness tuning, often treated primarily as a resonance-matching strategy, is shown here to act as a symmetry-control parameter that can switch nonlinear emission on or off without altering material composition or geometry. This observation generalizes naturally to multilayer stacks, guided-wave geometries, and other three-wave-mixing processes, where analogous symmetry-protected suppression of nonlinear coupling may arise even in strongly resonant systems.

Furthermore, the present results point to a conceptual refinement in the design of cavity-enhanced nonlinear optical systems. Maximizing the local field intensity or the linear quality factor alone is not sufficient for efficient frequency conversion. Nonlinear performance is fundamentally constrained by spatial-mode correlation and symmetry compatibility. The identification of nonlinearity-selective quasi-BICs therefore provides both a practical design principle for nonlinear photonic structures and a unifying framework for understanding why resonant nanophotonic systems can remain unexpectedly dark in the nonlinear regime.

\section{Acknowledgement}

AMK thanks the Asian Smart Cities Research Innovation Network for the scholarship, and NKG acknowledges generous funding support from BITS Pilani Hyderabad Campus and BITS Pilani GCIR under the NFSG scheme.

\bibliography{references}

\end{document}